\documentclass[runningheads]{llncs}
\usepackage{graphicx}
\usepackage{multirow}
\usepackage{array}
\usepackage{subcaption}
\usepackage{color, colortbl}
\usepackage{scrextend}
\definecolor{Gray}{gray}{0.9}
\definecolor{LightCyan}{rgb}{0.88,1,1}
\newcolumntype{g}{>{\columncolor{Gray}}p}
\begin{document}
\title{A comparative study of similarity-based and GNN-based link prediction approaches}
%
\titlerunning{Comparative study of similarity-based and GNN-based link prediction}
%
\author{Md Kamrul Islam \and
Sabeur Aridhi \and
Malika Smail-Tabbone}
\authorrunning{M. K. Islam et al.}
%
\institute{Universite de Lorraine, CNRS, Inria, LORIA, 54000 Nancy, France \\
\email{\{kamrul.islam, sabeur.aridhi, malika.smail\}@loria.fr}}
\maketitle              
\begin{abstract}
The task of inferring the missing links in a graph based on its current structure is referred to as link prediction. Link prediction methods that are based on pairwise node similarity are well-established approaches in the literature. They show good prediction performance in many real-world graphs though they are heuristics and lack of universal applicability. On the other hand, the success of neural networks for classification tasks in various domains leads researchers to study them in graphs. When a neural network can operate directly on the graph, then it is termed as the graph neural network (GNN). GNN is able to learn hidden features from graphs which can be used for link prediction task in graphs. Link predictions based on GNNs have gained much attention of researchers due to their convincing high performance in many real-world graphs. This appraisal paper studies some similarity and GNN-based link prediction approaches in the domain of homogeneous graphs that consists of a single type of (attributed)nodes and single type of pairwise links. We evaluate the studied approaches against several benchmark graphs with different properties from various domains. 


\end{abstract}
 
\section{Introduction}
One of the most interesting and long-standing problems in the field of graph mining is link prediction that predicts the probability of a link between two unconnected nodes based on available information in the current graph such as node attributes or graph structure~\cite{xu2016link}. The prediction of missing or potential links helps us toward the deep understanding of structure, evolution and functions of real-world complex graphs~\cite{shen2014reconstructing}. Some applications of link prediction include friend recommendation in social networks (\cite{adamic2003friends}), product recommendation in e-commerce~\cite{koren2009matrix}, knowledge graph completion~\cite{nickel2015review}, and finding interactions between proteins~\cite{airoldi2008mixed}.  

A large category of link prediction methods is based on some heuristics that measure the proximity between nodes to predict whether they are likely to have a link. Though these heuristics can predict links with high accuracy in many graphs, they lack universal applicability to different kinds of graphs. For example, the common neighbor heuristic assumes that two nodes are more likely to connect if they have many common neighbors. This assumption may be correct in social networks, but is shown to fail in protein-protein interaction (PPI) networks where two proteins sharing many common neighbors are actually less likely to interact~\cite{kovacs2019network}. In case of using these heuristics, it is required to manually choose different heuristics for different graphs based on prior beliefs or expensive trial and error process. On the other hand, learning-based link prediction approaches are able to learn suitable heuristics from the graph itself. The success of the neural network is well-known for machine learning task in many real-world applications like image classification~\cite{paoletti2018new}, speech recognition~\cite{hinton2012deep}, video processing~\cite{redmon2016you}, natural language processing~\cite{luong2015faster}. The applications can represent the data in Euclidean space and neural network is able to extract the hidden features from the data space. However, the neural network can not be applied directly into the graph domain due to two important challenges~\cite{wu2020comprehensive}. Firstly, a graph contains unordered nodes and a variable number of neighbours for each node. Secondly, the assumption of independence of data is no longer true for graphs as each node is linked to some other nodes. The first attempt to study the neural network in the graph domain was done in~\cite{scarselli2008graph}. Then, Graph Neural Networks (GNNs) has become a powerful tool for learning hidden features in graphs. In the last decades, researchers have developed many GNN-based methods which are used for several tasks completion such as graph classification~\cite{ying2018hierarchical}, node classification~\cite{kipf2016semi}, and link prediction~\cite{zhang2018link}.

In this paper, we first introduce the link prediction problem and highlight similarity-based and GNN-based methods. Then, we choose a few approaches from both link prediction categories to evaluate their performances on different types of graphs, namely simple or homogeneous graphs and node-attributed graphs. We compare their performance with respect to the prediction accuracy and computational time.

\section{Link Prediction: Problem and Approaches}
\begin{figure}
    \centering
    \begin{subfigure}[t]{0.5\textwidth}
        \centering
        \includegraphics[height=0.8in,width=1in]{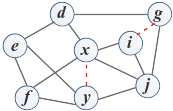}
        \caption{Graph at time t}
        \label{fig:lp1}
    \end{subfigure}%
    ~ 
    \begin{subfigure}[t]{0.5\textwidth}
        \centering
        \includegraphics[height=0.8in,width=1in]{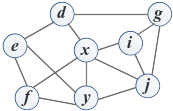}
        \caption{Graph at time $t + \Delta t$}
        \label{fig:lp2}
    \end{subfigure}
    \caption{Illustration of link prediction problem}
    \label{fig:lp}
\end{figure}

Consider an undirected graph at a particular time t where nodes represent entities and links represent the relationships between pair entities (or nodes). The link prediction problem is defined as discovering or inferring a set of missing links (existing but not observed) in the graph at time $t + \Delta t$. The problem can be illustrated with a simple undirected graph in Fig.~\ref{fig:lp}, where circles represent nodes and lines represent links between pair of nodes. Black solid lines represent observed links and red dashed lines represent missing links in the current graph. Fig.~\ref{fig:lp1} shows the snapshot of the graph at time t, where two missing links exist between node pairs (x, y) and (g, i). The link prediction problem aiming to predict the appearance of these two missing links as observed links in the graph in near future $t + \Delta t$, as illustrated in Fig.~\ref{fig:lp2}.

\subsection{Similarity-based Link Prediction}
The similarity-based approach is the most commonly used approach for link prediction which is developed based on the assumption that two nodes in a graph interact if they are similar. The definition of similarity is a crucial and non-trivial task that varies from domain to domain even from the graph to graph in the same domain~\cite{martinez2016survey}. As a result, numerous similarity-based approaches have been included in the literature to predict links in small to large graphs. Some similarity-based approaches use the local neighbourhood information to compute similarity score are known as local similarity-based approach. Another category of similarity-based approach is global approaches those use the global topological information of graph. The computational complexity of global approaches makes them unfeasible to be applied on large graphs as they use the global structural information such as adjacency matrix~\cite{martinez2016survey}. For this reason, we are considering only the local similarity-based approaches in the current study. We have studied 13 popular similarity-based approaches for link prediction. Table~\ref{Table:simappro} summarizes the approaches with the basic principle and similarity function. 


\renewcommand{\arraystretch}{0.8}
\begin{table}
\centering
\caption{Summary of studied similarity-based approaches. The similarity function is defined to predict a link between two nodes $x$ and $y$. $\Gamma x$ and $\Gamma x$ denote the neighbour sets of nodes $x$ and $y$ respectively. $r_{x,y}$ denotes the link between two nodes $x$, $y$.}
\label{Table:simappro}
\small
\begin{tabular}{|p{3cm}|p{5cm}|p{4cm}|}
\hline
\rowcolor{Gray}
\textbf{Approach} & \textbf{Principle}   & \textbf{Similarity-function} \\ 
\hline
Adamic-Adar (AA) ~\cite{adamic2003friends}    & Variation of CN where each common neighbour is logarithmically penalized by its degree      & $ S^{AA}(x, y) =\sum_{z\in \Gamma x \cap \Gamma y^{}} \frac{1}{log \left | \Gamma z \right |} $      \\ 
\hline
Common Neighbours (CN) ~\cite{lorrain1971structural}    & Two nodes are more likely to be linked share more neighbours     & $ S^{CN}(x, y) =\left | \Gamma x \cap \Gamma y  \right|$    \\ 
\hline
Resource Allocation (RA) ~\cite{zhou2009predicting}   & Based on the resource allocation process to further penalize the high degree common neighbours by more amount     & $ S^{RA}(x, y) =\sum_{z\in \Gamma x \cap \Gamma y^{}} \frac{1}{ \left | \Gamma z \right |} $       \\ 
\hline
Preferential Attachment (PA) ~\cite{barabasi1999emergence}    & Based on the rich-get-richer concept where the link probability between two high degree nodes is higher than two low degree nodes     & $ S^{PA}(x, y) = \left | \Gamma x \right | \times \left | \Gamma y \right | $      \\ 
\hline
Jaccard Index(JA) ~\cite{jaccard1901etude}       & Normalization of CN where the score is penalized for each non-common neighbour     & $ S^{JA}(x, y) = \frac{\left | \Gamma x \cap \Gamma y \right |}{\left | \Gamma x \cup \Gamma y \right |}  $     \\ 
\hline
Salton Index(SA) ~\cite{salton1983introduction}       & Motivated by cosine similarity that defines link probability based on cosine angle between adjacency vectors for nodes pair     & $ S^{SA}(x, y) = \frac{\left | \Gamma x \cap \Gamma y \right |}{\sqrt{\left | \Gamma x \right | \times \left | \Gamma y \right |}} $      \\ 
\hline
Sørensen Index(SO) ~\cite{sorensen1948method}    & Describing the overall proportion of common neighbours from a local perspective.     & $ S^{SO}(x, y) = \frac{2 \times \left | \Gamma x \cap \Gamma y \right |}{\left | \Gamma x \right | + \left | \Gamma y \right |} $     \\
\hline
Hub Promoted Index (HPI) ~\cite{ravasz2002hierarchical}     & Promoting link formation between high-degree nodes and hubs    & $ S^{HPI}(x, y) = \frac{\left | \Gamma x \cap \Gamma y \right |}{max(\left | \Gamma x \right |, \left | \Gamma y \right |)}  $    \\
\hline
Hub Depressed Index (HDI) ~\cite{ravasz2002hierarchical}    & Promoting link formation between low-degree nodes and hubs.    & $ S^{HDI}(x, y) = \frac{\left | \Gamma x \cap \Gamma y \right |}{min(\left | \Gamma x \right |, \left | \Gamma y \right |)}  $   \\
\hline
Local Leicht-Holme- Newman (LLHN) ~\cite{leicht2006vertex} & Utilizing both of real and expected amount of common neighbours between a pair of nodes to define their similarity. & $ S^{LLHN}(x, y) = \frac{\left | \Gamma x \cap \Gamma y \right |}{\left | \Gamma x \right | \times \left | \Gamma y \right |}  $\\
\hline
Individual Attraction (IA) ~\cite{dong2011link} & Maximizing the likelihood of link formation for highly interlinked nodes pair.   & $ S^{IA}(x, y) =\sum_{z\in \Gamma x \cap \Gamma y^{}} \frac{\left | r_{z,\Gamma x \cap \Gamma y} \right |+2}{ \left | \Gamma z \right |} $ \\
\hline
Cannistrai–Alanis– Ravai (CAR) ~\cite{cannistraci2013link} & Utilization of level-2 links along with common neighbourhood information in computing the pairwise similarity score   & $ S^{CAR}(x, y) =\sum_{z\in \Gamma x \cap \Gamma y} 1+ \frac{\left | \Gamma x \cap \Gamma y \cap \Gamma z \right |}{2} $ \\
\hline
Clustering Coefficient-based Link Prediction (CCLP) ~\cite{wu2016link}   & Quantification of the contribution of each common neighbour by utilizing the local clustering coefficient of the node.  & $ S^{CCLP}(x, y) =\sum_{z\in \Gamma x \cap \Gamma y^{}} CC_{z} $       \\ 
\hline
\end{tabular}
\end{table}

These approaches except CCLP use node degree, common neighborhood or links among common neighborhood information to compute similarity scores. CCLP uses the clustering coefficient (CC) of each common neighbour to compute the role of its to the similarity score. The clustering coefficient is defined as the ratio of the number of triangles and the expected number of triangles passing through a node. If $t_z$ is the number of triangles passing through node z and $\Gamma z$ is the neighbourhood of z then the clustering coefficient ($CC_z$) of node z is defined as 
\begin{equation}
    CC_{z}=\frac{2\times t_z }{\left | \Gamma z \right | (\left |\Gamma z \right | -1) }
\end{equation}
Overall, these local similarity-based approaches except PA work well when the graphs have a high number of common neighbours between a pair of nodes. However, the SA, HDI and LLHN suffer from outlier when one of the two nodes has no neighbour. In addition, some of the approaches like JA, SO, HPI suffer from the outlier when both of the nodes have no neighbour. 

\subsection{Graph Neural Network(GNN)-based Link Prediction}
Graph neural network (GNN) is an extension of the neural network to be applied to graph data. A GNN computes the node representation based on the available node  information. It aggregates the information from its neighbours to find its final representation and the representation is fed into a multi-layer neural network for several downstream tasks like node classification, link prediction and graph classification. Based on the architecture, GNNs are broadly categorized into five categories: recurrent graph neural network (RecGNN), convolution graph neural network (ConvGNN), graph auto-encoder (GAE), and spatial-temporal graph neural network (STGNN)~\cite{wu2020comprehensive}. RecGNNs are the pioneers of GNNs those work based on the assumption that the nodes constantly exchange the information with the neighbours until a stable state is reached. Motivated by the convolution operation of the neural network in the image domain, ConvGNNs compute the embedding of a node by aggregating its own information and neighbours information. GAEs are the unsupervised version of GNN those encode the nodes into a latent vector space and reconstruct the graph to learn the embedding. STGNNs are used to learn the hidden features in a spatio-temporal graph based on the spatial and temporal dependency with time. Recently, researchers have studied the attention mechanism in RecGNN and ConvGNN to improve the prediction performance by allowing them to focus on the most relevant parts of the graph~\cite{velivckovic2018graph}. ConvGNNs has become popular in recent years due to its efficient graph convolution operation~\cite{wu2020comprehensive,zhang2020deep}. In this paper, we focus on the link prediction approaches based on ConvGNN. A ConvGNN starts with defining the neighbourhood $\Gamma v_i$ of each node $v_i$ in the graph $G(V,E)$ which is a crucial task as it can affect the accuracy and computational time. Some popular neighbourhood definitions include immediate neighbours, multi-hop neighbours~\cite{zhang2018link,xu2018representation}, sampling-based neighbours~\cite{zhang2017weisfeiler,huang2018adaptive}. The feature vector, $x_i$ of each node, $v_i$ is then computed based on its attribute and structural information. The feature vectors of nodes are fed into a stack of layers to learn the hidden features of the graph. A simple ConvGNN update the node representation in each layer in the following three basic steps~\cite{zhang2020deep,ying2019gnnexplainer}
\begin{enumerate}
    \item \textbf{Computation of neural messages:} The neural messages of each link for next layer is computed based on the current representations of both end nodes of the link. If $h_{i}^{l}$ and $h_{j}^{l}$ are current representations of a nodes pair ($v_i, v_j$), the message of the link is defined as 
    \begin{equation}
        m_{ij}^{l+1}=MSG(h_{i}^l,h_{j}^l,r_{ij})
    \end{equation}
    Here, $l$ represents the current layer, $r_{ij}\in E$ is the relation between the nodes pair and $MSG$ is the message computation function for the links. Many GNN models use the link types ~\cite{zitnik2018modelling} or link weight \cite{ying2018graph} for encoding $r_{ij}$. The initial representations of nodes $v_i$ and $v_j$ are $x_i$ and $x_j$ respectively (i.e. $h_{i}^{1}=x_i$ and $h_{j}^{1}=x_j$).
    \item \textbf{Aggregating the neighbour information:} The next operation of the layer is to aggregate the neighbour messages $m_{ij}^l$ for node $v_i$. An aggregation function is defined as 
    \begin{equation}
        M_{i}^{l+1}=AGGR({m_{ij}^{l+1} | v_j\in\Gamma v_i })
    \end{equation}
    Here, $\Gamma v_i$ is the set of neighbours of node $v_i$ and $AGGR$ is the aggregation function. Some popular aggregation functions exist in the literature such as mean/max pooling ~\cite{hamilton2017inductive}, sort pooling~\cite{zhang2018end}, permutation invariant~\cite{xu2018powerful}.
    \item \textbf{Updating the node representation:} In this step, the representation or embedding of node $v_i$ in the next layer is updated based on the current embedding, $h_i^l$ and the aggregated message, $M_{i}^{l+1}$
        \begin{equation}
        h_{i}^{l+1}=UPDATE(h_{i}^{l},M_{i}^{l+1})
    \end{equation}
    Here, the $UPDATE$ function is a non-linear function like sigmoid, recified linear unit(ReLU), hyperbolic tangent(TanH). The output embedding $h_i^{l+1}$ is the input for next layer.
\end{enumerate}
Each layer in the model follows these three steps and generates nodes embedding. The embedding from the last layer is fed into a standard classifier such as multilayer perception (MLP) with a softmax layer for downstream tasks. The parameters of the classifier are optimized using optimizer like Adam, stochastic gradient descent(SGD) along with loss functions such as cross-entropy, mean absolute error(MAE), mean squared error(MSE) and backpropagation.

There exist many link prediction approaches based on ConvGNN in the literature. Most of them are applicable to homogeneous graphs and few of them are applicable to heterogeneous graphs which consist of multiple types of nodes and links, node and link attributes and multiple links between pairs of nodes. We study two recent GNN-based link prediction approaches which are applicable to homogeneous graphs only as our study is confined to those graphs. The first one is WLNM (Weisfeiler-Lehman Neural Machine) that utilizes only the structural information of nodes for the link prediction task. SEAL (Sub-graphs, Embeddings and Attributes) is the second one that uses the structural, latent and attribute information of node for the same task. The approaches are briefly described below.
\subsubsection{Weisfeiler-Lehman Neural Machine (WLNM)}
\begin{figure}
    \centering
    \includegraphics[width=0.8\linewidth]{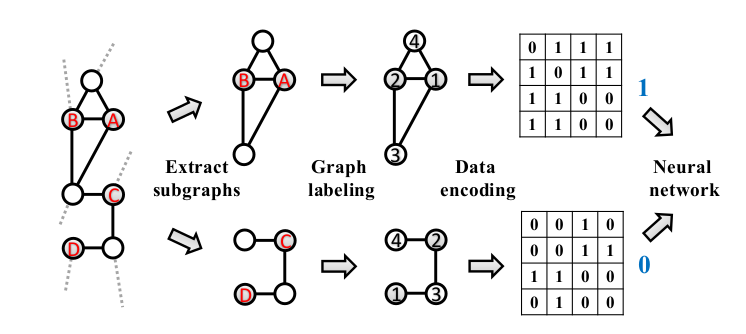}
    \caption{Illustration of WLNM approach~\cite{zhang2017weisfeiler}}
    \label{fig:wlnm}
\end{figure}

Based on the well-known Weisfeiler-Lehman canonical labelling algorithm~\cite{weisfeiler1968reduction}, Zhang \&Chen developed a link prediction approach for graph called Weisfeiler-Lehman Neural Machine (WLNM)~\cite{zhang2017weisfeiler}. WLNM learns the structural features from the graph and uses in prediction task. WLNM is a three steps link prediction approach that starts with extracting sub-graphs, labelling and encoding the nodes and ends with training and evaluating the neural network. Fig.~\ref{fig:wlnm} illustrates the training process of WLNM with one existent link (A,B) and one non-existent link(c,d).
The three steps of WLNM link prediction approach are described as following.
\begin{enumerate}
    \item \textbf{Sub-graph extraction:} WLNM starts with extracting the k-vertex neighbouring sub-graph of a link called enclosing sub-graph. k is the user-defined parameter that defines the size of sub-graph. For a given link, 1-hop neighbours are added in the sub-graph, then 2-hop neighbours and so on until the number of neighbours is greater or equal to k. If there are k' nodes in sub-graph such that $k'>k$ then $k'-k$ nodes with higher hop number are removed sub-graph.
    \item \textbf{Node labelling and encoding:} Weisfeiler-Lehman (WL) is a popular graph labelling algorithm that uses the concept of signature string for each node to compute node labels~\cite{weisfeiler1968reduction}. Instead of using classical WL algorithm, WLNM develops a hashing based color refinement process for faster node labelling. If two nodes have still the same label, WLNM uses the naughty node labelling algorithm to break the tie~\cite{mckay2014practical}. The nodes are sorted according to the node label in increasing order and an upper triangular adjacency matrix is computed.
    \item \textbf{Neural network training and evaluation:} WLNM uses a fully connected multi-layer perception (MLP) neural network to learn structural features from the sub-graph. The output layer of the MLP is a softmax layer that classifies the link into two classes (existent and non-existent). The upper triangular adjacency matrix of the sub-graph is vertically fed into the MLP to train and evaluate WLNM approach. The neural network is trained for both of existent and non-existent links.
\end{enumerate}
WLNM is a simple GNN-based link prediction approach which is able to learn the link prediction heuristics from a graph. In contrast to similarity-based heuristics, WLNM has universal applicability properties.  However, WLNM truncates some neighbours to limit the number of nodes in the sub-graph to a user-defined size. The truncated neighbours may be informative for the prediction task.

\subsubsection{Learning from Sub-graphs, Embeddings and Attributes (SEAL)}
\begin{figure}
    \centering
    \includegraphics[width=\linewidth]{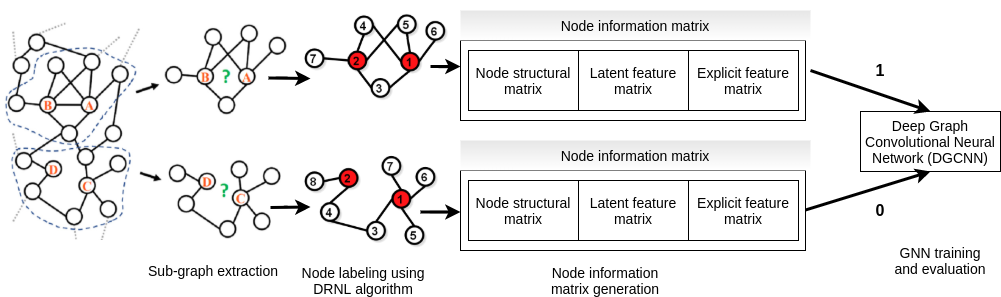}
    \caption{Architecture of SEAL approach}
    \label{fig:seal}
\end{figure}

Zhang et al~\cite{zhang2018link} developed a ConvGNN-based link prediction approach namely SEAL to learn from latent and explicit features of nodes along with the structural information of graph. Unlike WLNM, SEAL is able to handle neighbours of variable size. SEAL replaces the fully-connected neural network in WLNM with a graph neural network to learn the graph features efficiently. The overall architecture of the approach is shown in Fig.~\ref{fig:seal}. 
Like WLNM, SEAL also consists of three major steps which are described as follows: 
\begin{enumerate}
    \item \textbf{Sub-graph extraction and node labelling:} Likewise WLNM, SEAL approach uses the concept of local sub-graph instead of the whole graph for a link in prediction task. SEAL defines the sub-graph as the h-hop neighbours of a link which is built by the union operation on the h-hop neighbours of nodes of the link. For example, a 1-hop enclosing sub-graph contains all first-order or immediate neighbours, a 2-hop enclosing sub-graph contains all first-order and second-order neighbours. In Fig.~\ref{fig:seal}, the sub-graph for link (A,B) consists of 7 nodes(5 neighbours and 2 end nodes) and the sub-graph for link (C,D) consists of 8 nodes(6 neighbours and 2 end nodes). The approach shows that setting a small h can still provide good prediction performance. Then a unique label is assigned to each node of the sub-graph to indicate its importance in the prediction task. SEAL designs a new node labelling algorithm namely DRNL (double-radius node labelling) based on the topological distances of the node from both ends of the link in the sub-graph.
    \item \textbf{Node information matrix construction:} The information matrix of a node in SEAL is defined based on its structural label(structural feature), embedding(latent feature) and attribute(explicit feature). One hot encoding technique is applied to the labelled sub-graph to compute the structural vector of nodes. The structural feature vector of the node is then concatenated with the latent feature vector of the node. The latent feature is the low-dimensional latent representation/embedding of a node which is obtained by factorizing the adjacency matrix from the graph. SEAL uses the Node2Vec~\cite{grover2016node2vec} algorithms to learn the latent feature vector for each node in sub-graph. The last part of the information vector of the node is an explicit feature vector which is computed based on the continuous or discrete attributes of the node. One hot coding technique is used to find the explicit feature vector of each node.
    \item \textbf{Neural network training and evaluation:} The learned node information matrix of the sub-graph is feed into a GNN called DGCNN (Deep Graph Convolutional Neural Network)~\cite{zhang2018end} to perform the link prediction task. DGCNN consists of propagation-based convolution layer and aggregation layer to aggregate the neighbour's information vector. DGCNN uses a sort-pooling layer to unify the size of the representation of the sub-graph.
\end{enumerate}
SEAL utilizes the available information in the graph to improve the prediction performance. However, SEAL is limited to be applied on homogeneous graphs though many real work graphs are heterogeneous graphs. Moreover, the use of latent feature affects the computational time of SEAL.

\section{Experimental Design}

\subsection{Datasets Characteristics}
We perform the comparative study of the above discussed similarity and GNN based link prediction approaches in graphs from different domains. To evaluate and describe the performance of the link prediction approaches, we choose ten benchmark graphs from different areas: Ecoli~\cite{salgado2001regulondb}, FB15K~\cite{bordes2013translating}, NS~\cite{newman2006finding}, PB~\cite{ackland2005mapping},Power~\cite{watts1998collective}, Router~\cite{spring2002measuring}, USAir~\cite{handcock2003statnet}, WN18~\cite{bordes2014semantic}, YAGO3-10~\cite{mahdisoltani2013yago3}, and Yeast~\cite{von2002comparative}. Ecoli and Yeast are two biological graphs those represent the biological relations between operons in Escherichia Coli bacteria and protein-protein interaction in yeast. PB (Political Blog) graph represents the network among political blog pages in US where the blog pages are identified as nodes and hyperlinks between the blog pages are identified as links of the graph. We consider the original directed links as undirected links. Net Science (NS) graph represents a collaboration network of researchers who publish papers on network science. Power is an electrical grid network of western US representing the network describing high voltage transmission among generators, transformers and substations. The Router graph represents the router-level internet where each router has an identifier and undirected links with other routers. The USAir graph represents the network of the US air transportation system that consists of attributed nodes (airports) and links between two airports. FB1K, WN18 and YAGO3-10 are simplified knowledge graphs. The original FB15K is a Freebase Knowledge Graph which was extracted from Wikidata and DBPedia. This knowledge graph contains 540188 triples where each triple consists of identifiers of freebase entity with a relationship name between them. WN18 is another knowledge graph that is a large lexical graph of English. The last knowledge graph is YAGO3-10 that was prepared at the Max Planck Institute for Computer Science in Saarbrucken in 2015. These knowledge graphs consist of subject-relationship type-object triples. However, as most studied approaches are applicable to homogeneous graphs only, 
we simplify these knowledge graphs by overlooking the types of relationships and reducing multiple links to single links between nodes/entities. All of the graphs are considered as undirected graphs. In this study, we are considering them as large graphs instead of knowledge graphs.

We use the Gephi tool~\cite{bastian2017gephi} to extract the topological statistics of the graphs. The characteristics of the graph datasets are summarized in Table~\ref{stTable}. Based on the number of nodes, these graphs are categorized into small/medium graphs with less or equal 10000 nodes and large graphs with more than 10000 nodes. 


\renewcommand{\arraystretch}{1}
\begin{table}
\centering
\caption{Topological statistics of graph datasets: number of nodes(\#Node), links(\#Link), average node degree (NDeg), total triangle(\#Triangle), clustering coefficient (C.Coef), average path length (APL), network diameter (Diam) and type of graph.}
\begin{tabular}{|l|l|l|l|l|l|l|l|l|l|l|l|} 
\hline
\rowcolor{Gray}
Graphs & \#Node   & \#Link    & NDeg & \#Triangle  & C.Coef    & APL  & Diam & Graph type  \\ 
\hline
Ecoli    & 1805     & 42325     & 46.898    & 459809    & 0.350     & 2.714     & 10 & Homogeneous    \\
\hline
FB15K    & 14949    & 260183    & 44.222    & 565104    & 0.218     & 2.716     & 8 & Homogeneous    \\
\hline
NS       & 1461     & 2742      & 3.754     & 3764      & 0.878     & 5.823     & 17 & Homogeneous    \\
\hline
PB       & 1222     & 14407     & 23.579    & 49549     & 0.239     & 2.787     & 8  & Homogeneous    \\ 
\hline
Power    & 4941     & 6594      & 2.669     & 651       & 0.107     & 18.989    & 46 & Homogeneous    \\ 
\hline
Router   & 5022     & 6258      & 2.492     & 803       & 0.033     & 6.449     & 15 & Homogeneous    \\ 
\hline
USAir    & 332      & 2126      & 12.807    & 12181     & 0.749     & 2.738     & 6 & Homogeneous,    \\ 
    &       &       &     &      &      &     &  & node-attributed     \\ 
\hline
WN18     & 40943    & 75769     & 3.709     & 5107      & 0.077     & 7.426     & 18 & Homogeneous    \\
\hline
YAGO3-10 & 113273   & 758225   & 18.046    & 225094    & 0.114     &  22.999   &   14 & Homogeneous   \\
\hline
Yeast    & 2375     & 11693     & 9.847     & 60689     & 0.388     & 5.096     & 15 & Homogeneous    \\ 
\hline
\end{tabular}
\label{stTable}
\end{table}


\subsection{Construction of Train and Test sets}

We follow a random sampling validation protocol to evaluate the performance of the studied  approaches~\cite{zhang2017weisfeiler,wang2019dgl}. The train and test datasets are prepared from a graph G(V, E), where V is the set of vertices and E is the set of existent links. Two types of both training and test datasets are prepared from the graph. The first training dataset is positive training dataset that contains randomly selected 90\% observed links and an equal number of non-existent links form the negative training dataset. The remaining 10\% existent links form the positive test dataset and an equal number of non-existent links form the negative test dataset. At the same time, the graph connectivity of the training set and the test set is guaranteed. We prepare five train and five test datasets for evaluating the performance of the approaches.


For evaluating the performance of similarity-based approaches, the graph is built from the positive training dataset whereas, for graph neural network-based approaches, the graph is built from the original graph that contains both of positive train and test datasets. However, a link is temporarily removed from the graph to train it to the GNN-based approaches or to predict its existence. The approaches are evaluated on positive and negative test datasets. For WLNM, we choose the neighbour size to 10 and for SEAL we choose the hop to 1 for all graphs. The similarity scores of similarity-based approaches for test links are computed based on the training graphs which contain only train links. The performance of link prediction approach is quantified by defining two standard evaluation metrics, precision and AUC (Area Under the Curve). All of the approaches are run on a Dell Latitude 5400 machine with 32GB primary memory and core i7 (CPU 1.90GHz) processor. 

\subsection{Computation procedures for Precision and AUC}
Precision describes the fraction of missing links that are accurately predicted as existent link~\cite{yang2016predicting,pan2016predicting,wu2018improving}. To compute the precision, all of the predicted links from a test set are ranked in decreasing order of their scores. If $L_r$ is the number of existing links (in the positive test set) among the L-top ranked predicted links then the precision is defined as 
\begin{equation}
    Precision=\frac{L_r}{L}
    \label{eq:pr}
\end{equation}
The precision is a measure of result relevance. The higher the precision indicates the higher accuracy of the prediction approach. An ideal prediction approach has a precision of 1.0 that means all the missing links are accurately predicted. We set L to the number of existent links in the test set.

On the other hand, the metric AUC is measured to demonstrate the ability of an approach in distinguishing between an existent and a non-existent link. 
It is defined as the probability that a randomly chosen missing link has a higher similarity score than a randomly chosen non-existent link~\cite{pan2016predicting}. Suppose, n existent and n non-existent links are chosen from positive and negative test sets. If $n_1$ is the number of existent links having a higher score than non-existent links and $n_2$ is the number of existent links having equal score as non-existent links then AUC is defined as 
\begin{equation}
    AUC=\frac{n_1+0.5n_2}{n}
    \label{eq:auc}
\end{equation}
An AUC of more than 0.5 indicates that the prediction index has a better effect than choosing links randomly and vice versa. Generally, the degree to which AUC exceeds 0.5 indicates how much good the prediction approach. We consider half of the total links in the positive test set and negative test set to compute AUC. 

\section{Analysis of the Results}

\subsection{Comparison of Prediction Accuracy with Precision and AUC}
The prediction approaches are evaluated in each of the five sets (train and test set) of each graph and performance metrics (precision, AUC) are recorded. The maximum and minimum similarity scores are computed from the top-L for each test set of each graph. Table~\ref{Table:small} shows the mean maximum(Max Score) and minimum similarity (Min Score) scores for each similarity-based in each graph. We measure the precision in two different ways based on the top-L test links. Firstly, we use Equation~\ref{eq:pr} as it is where $L_r$ is the number of positive links in top-L test links. However, the minimum similarity scores for many similarity-based approaches are very low (close to 0) that creates  difficulty to make a separation between some positive and negative test links. To overcome this problem, we define a threshold when defining $L_r$. However, defining threshold to similarity-based approaches is again a non-trivial task as the maximum and minimum scores vary for different graphs and even for different test sets. To overcome this problem, we define a threshold as the average of the maximum and minimum score in top-L links. We compute the number of positive test links in top-L links (as $L_r$) as those having similarity scores above the threshold. We compute the threshold-based precision only for similarity-based approaches as GNN-based approaches do learn the threshold. The corrected precision is shown in parentheses in Table~\ref{Table:small}. Each value of the table is the mean over the five test sets. The evaluation metrics precision and AUC for the studied approaches in the seven small to medium-size graphs are tabulated in Table~\ref{Table:small}.

Table~\ref{Table:small} shows that, overall, the similarity-based approaches give high precision (without defining threshold) and AUC values in well-connected (high clustering coefficient, high node degree) graphs while GNN-based approaches show good precision and AUC in all graphs. In the Ecoli graph, CCLP shows the highest precision (0.96) while the lowest precision(0.78) is recorded for the PA approach. The precisions of other similarity-based approaches are close to the highest precision score. The highest clustering coefficient contributes to the success of CCLP in terms of precision in Ecoli. However, the precision of similarity-based approaches drops drastically when computing precision based on the threshold as many positive links with very low similarity scores (even 0) comparing to the threshold. The precision of WLNM and SEAL approaches are lower than the similarity-based approaches and they are 0.867 and 0.807 respectively.
\renewcommand{\arraystretch}{1}
\begin{table}
\centering
\caption{AUC and Precision values with Max Scores and Min Scores in small/medium graphs. Precision in () is computed based on threshold in top-L links. Graph-wise highest/lowest metrics are indicated in bold fonts while approach-wise highest/lowest metrics are shown in italic.}
\label{Table:small}
\scriptsize
\begin{tabular}{|p{1.3cm}|g{1.2cm}|g{1.2cm}|g{1.2cm}|g{1.2cm}|g{1.2cm}|g{1.2cm}|g{1.2cm}|g{1.2cm}|}
\hline
\rowcolor{Gray}
\textbf{App.}   & \textbf{Metrics}   & \textbf{Ecoli} & \textbf{NS} & \textbf{PB} & \textbf{Power} & \textbf{Router} & \textbf{USAir} & \textbf{Yeast} \\ \hline
\rowcolor{white}
\multirow{3}{*}{AA} & Precision &0.90(0.06)   &0.87(0.15)   &\textbf{0.86}(0.01)   &0.17(0.02)   &\textit{\textbf{0.07}}(0.01)  &\textit{0.92}(0.16)   &0.83(0.06)   \\ \cline{2-9} 
            & Max scor   &32.84  &5.83   &33.41  &3.04   &5.60   &16.69  &23.71   \\ \cline{2-9} 
            \rowcolor{white}
            & Min scor   &2.86   &1.14   &0.58   &0.00   &0.00   &2.70   &0.00    \\ \cline{2-9} 
            & AUC       &0.93   &\textit{0.94}   &0.92   &0.58   &\textit{0.54}   &\textit{0.94}   &0.91     \\ \hline
\rowcolor{white}
\multirow{3}{*}{CN} & Precision &0.91(0.07)   &0.87(0.22)   &\textbf{0.86}(0.02)   &0.17(0.04)   &\textit{\textbf{0.07}}(.004)  &\textit{0.92}(0.23)   &0.83(0.06)   \\ \cline{2-9} 
            & Max scor   &153    &11.0   &119    &29.0   &15.0   &51.0   &90.33   \\ \cline{2-9} 
            \rowcolor{white}
            & Min scor   &12.0   &1.40   &3.00   &0.00   &0.00   &9.67   &0.00  \\ \cline{2-9}
            & AUC       &0.93   &0.93   &0.91   &0.58   &\textit{0.54}   &\textit{0.95}   &0.91   \\ \hline 
\rowcolor{white}
\multirow{3}{*}{PA} & Precision &\textbf{0.78}(0.05)  &\textbf{0.69}(0.02)  &0.83(0.01)   &0.49(0.02)   &\textit{0.41}(0.01)   &\textit{\textbf{0.85}}(0.13)   &\textbf{0.79}(0.06)   \\ \cline{2-9} 
            & Max scor   & 65679  &362.0  &61052  &53.0   &2397   &8298.7 &10642   \\ \cline{2-9} 
            \rowcolor{white}
            & Min scor   &3532   &12.0   &855.7  &4.0    &1.0    &739.3  &95.0  \\ \cline{2-9}
            & AUC       &\textbf{0.80}  &\textbf{0.66}  &\textit{0.90}   &\textbf{0.46}  &\textit{\textbf{0.43}}  &\textit{0.90}   &\textbf{0.86} \\ \hline
\rowcolor{white}
\multirow{3}{*}{RA} & Precision &0.91(0.03)   &0.87(0.15)   &\textbf{0.86}(0.01)  &0.17(0.03)   &\textit{\textbf{0.07}}(0.01)  &\textit{0.92}(0.10)   &0.83(0.07)   \\ \cline{2-9} 
            & Max scor   &1.70   &1.80   &4.19   &0.84   &1.32   &2.83   &2.37   \\ \cline{2-9}
            \rowcolor{white}
            & Min scor   &0.19   &0.40   &0.03   &0.00   &0.00   &0.32   &0.00   \\ \cline{2-9}
            & AUC       &\textit{0.94}   &\textit{0.94}   &0.92   &0.58   &0.54   &\textit{0.94}   &0.91   \\\hline  
\rowcolor{white}
\multirow{3}{*}{JA} & Precision &\textit{0.90}(0.11)   &0.87(0.42)   &0.79(0.07)   &0.17(0.07)   &\textit{\textbf{0.07}}(0.01)  &0.88(0.18)   &0.83(0.34)   \\ \cline{2-9} 
            & Max scor   &0.49   &0.60   &0.37   &0.60   &0.39   &0.45   &0.50   \\ \cline{2-9}
            \rowcolor{white}
            & Min scor   &0.10   &0.09   &0.04   &0.00   &0.00   &0.17   &0.00  \\ \cline{2-9} 
            & AUC       &\textit{0.94}   &0.92   &0.87   &0.58   &\textit{0.53}   &0.92   &0.91   \\ \hline
\rowcolor{white}
\multirow{3}{*}{SA} & Precision &\textit{0.91}(0.10)   &0.87(0.67)   &0.80(0.06)   &0.17(0.07)   &\textit{\textbf{0.07}}(0.01)  &0.90(0.15)   &0.83(0.40)   \\ \cline{2-9}
            & Max scor   &0.98   &1.00   &0.75   &0.94   &0.90   &0.91   &1.00   \\ \cline{2-9}
            \rowcolor{white}
            & Min scor   &0.22   &0.51   &0.11   &0.11   &0.00   &0.41   &0.00  \\ \cline{2-9}
            & AUC       &\textit{0.94}   &\textit{0.94}   &0.87   &0.57   &\textit{0.54}   &0.91   &0.90      \\ \hline 
            
\rowcolor{white}
\multirow{3}{*}{SO} & Precision &\textit{0.90}(0.11)   &0.87(0.64)   &0.79(0.07)   &0.17(0.06)   &\textit{\textbf{0.07}}(0.01)  &0.88(0.18)   &0.83(0.34)   \\ \cline{2-9}
            & Max scor   &0.98   &1.00   &0.74   &0.93  &0.90    &0.91   &1.00   \\ \cline{2-9}
            \rowcolor{white}
            & Min scor   &0.19   &0.46   &0.07   &0.00  &0.00    &0.34   &0.00  \\ \cline{2-9} 
            & AUC       &\textit{0.94}   &\textit{0.94}   &0.87   &0.57   &\textit{0.54}   &0.90   &0.91   \\ \hline
\rowcolor{white}
\multirow{3}{*}{HPI}& Precision &0.90(0.20)   &0.87(\textbf{0.96})   &0.80(0.15)   &0.17(0.13)   &\textit{\textbf{0.07}}(0.02)  &\textit{0.91}(0.45)   &0.83(0.70)   \\ \cline{2-9} 
            & Max scor   &1.00   &1.00   &1.00   &1.00   &1.00   &1.00   &1.00   \\ \cline{2-9} 
            \rowcolor{white}
            & Min scor   &0.33   &0.83   &0.21   &0.00   &0.00   &0.77   &0.00  \\ \cline{2-9}
            & AUC       &\textit{0.94}   &\textit{0.94}   &0.85   &0.58   &\textit{0.54}   &0.91   &0.90   \\ \hline
\rowcolor{white}
\multirow{3}{*}{HDI} & Precision &\textit{0.90}(0.08)  &0.87(0.64)   &0.79(0.05)   &0.17(0.03)   &\textit{\textbf{0.07}}(0.01)  &0.88(0.18)   &0.83(0.24)   \\ \cline{2-9}
            & Max scor   &0.97   &1.00   &0.68   &0.89   &0.89   &0.85   &1.00   \\ \cline{2-9} 
            \rowcolor{white}            
            & Min scor   &0.14   &0.33   &0.05   &0.00   &0.00   &0.24   &0.00  \\ \cline{2-9}
            & AUC       &\textit{0.94}   &\textit{0.94}   &0.86   &0.58   &\textit{0.53}   &0.90   &0.91       \\ \hline
\rowcolor{white}
\multirow{3}{*}{LLHN} & Precision &\textit{0.89}(.001) &0.87(0.13)   &\textbf{0.74}(.001)  &0.17(0.03)   &\textit{\textbf{0.07}}(.003)  &0.87(0.03)   &0.83(0.01)   \\ \cline{2-9} 
            & Max scor   &0.32   &1.00   &0.42   &2.06   &0.83   &0.58   &0.67   \\ \cline{2-9} 
            \rowcolor{white}
            & Min scor   &0.00   &0.10   &0.00   &0.00   &0.00   &0.01   &0.00  \\ \cline{2-9} 
            & AUC       &0.91   &\textit{0.93}   &\textbf{0.76}  &0.58   &\textit{0.53}   &\textbf{0.77}  &0.90   \\ \hline
\rowcolor{white}
\multirow{3}{*}{IA} & Precision &0.90(0.07)   &0.87(0.27)   &0.85(0.03)   &0.17(0.12)   &\textit{\textbf{0.07}}(0.01)  &\textit{0.92}(0.28)   &0.83(0.07)   \\ \cline{2-9} 
            & Max scor   &149.4  &10.7   &91.2   &4.3    &8.1    &46.6   &80.6  \\ \cline{2-9}
            \rowcolor{white}
            & Min scor   &12.5   &2.6    &3.5    &0.0    &0.0    &11.2   &0.0    \\ \cline{2-9} 
            & AUC       &\textit{0.93}   &\textit{0.93}   &0.91   &0.58   &\textit{0.54}   &\textit{0.93}   &0.91   \\ \hline
\rowcolor{white}
\multirow{3}{*}{CAR} & Precision &0.91(0.04)  &0.87(0.18)   &\textbf{0.86}(0.02)   &0.17(0.03)   &\textit{\textbf{0.07}}(0.01)  &\textit{0.92}(0.24)   &0.83(0.06)   \\ \cline{2-9} 
            & Max scor   &4833   &46.0   &1515.2 &2.3    &25.2   &555    &1831   \\ \cline{2-9} 
            \rowcolor{white}
            & Min scor   &50.2   &1.4    &3.0    &0.0    &0.0    &46.0   &0.0    \\ \cline{2-9}
            & AUC       &\textit{0.93}   &\textit{0.93}   &0.91   &0.59   &\textit{0.54}   &0.91   &0.91   \\\hline  
\rowcolor{white}
\multirow{3}{*}{CCLP} & Precision   &\textit{\textbf{0.96}}(0.06)  &0.73(0.21)   &\textbf{0.86}(0.01)   &\textbf{0.08}(0.01)  &\textit{\textbf{0.07}}(0.01)   &0.91(0.18)   &0.82(0.06)   \\ \cline{2-9} 
            & Max scor   &30.6   &8.0    &27.0   &1.2    &1.1    &21.1   &39.2   \\ \cline{2-9} 
            \rowcolor{white}
            & Min scor   &1.8    &0.3    &0.3    &0.0    &0.0    &2.9    &0.0    \\ \cline{2-9}
            & AUC       &\textit{\textbf{0.95}}   &0.87   &0.91   &0.54   &\textit{0.53}   &0.94   &0.90   \\ \hline 
\rowcolor{white}
\multirow{3}{*}{WLNM} & Precision &0.87 &0.84   &\textit{0.78}   &\textbf{0.84}  &\textit{\textbf{0.89}}  &0.85   &0.87   \\ \cline{2-9} 
            & AUC       &0.93   &\textit{0.95}   &0.93   &\textit{0.76}   &0.92   &0.86   &0.86   \\ \hline
\rowcolor{white}
\multirow{3}{*}{SEAL} & Precision &0.81 &\textit{\textbf{0.96}}  &0.80   &\textit{0.66}   &0.80   &\textbf{0.94}  &\textbf{0.89}  \\ \cline{2-9} 
            & AUC    &\textbf{0.95} &\textit{\textbf{0.99}}  &\textbf{0.94}  &\textit{\textbf{0.77}}  &\textbf{0.94}  &\textbf{0.96}  &\textbf{0.98}  \\ \hline
\end{tabular}
\end{table}
The highest and lowest AUC values in Ecoli are found for SEAL and PA approaches respectively. The AUC value of another GNN-based approach WLNM is also very high and close to the highest AUC value. The high values of these two GNN-based approaches state that they are highly efficient in distinguishing between existent and non-existent links in Ecoli graph. Similar performance is found for other well-connected graphs (NS, PB, USAir and Yeast). In NS graph, SEAL performs with the best precision (0.96) and AUC (0.99) score and PA is the worst approach which shows the lowest precision and AUC values of 0.69 and 0.66 respectively. The precision scores of other approaches lie between 0.8 to 0.9 while the AUC values are between 0.9 to 0.95. A remarkable precision(highest) is found for HPI in NS graph while the precision scores of some similarity-based approaches like AA, CN, PA, RA are still very low when applying the threshold method. Overall, the AUC values of GNN-based approaches are higher than the similarity-based approaches in NS graph. In PB graph, the highest precision score is recorded in similarity-based approaches RA, CAR and CCLP whereas the highest AUC value is found for the GNN-based approach SEAL. LLHN performs worst in PB concerning both metrics. The precision of other approaches near or above 0.8. The high average node degree plays a role in most of the similarity-based approaches in performing better than the GNN-based approaches in terms of precision scores in PB graph. However, the precision of similarity-based approaches drops to below 0.2 when applying the threshold in computing precision. Similarity-based approaches shows very low precisions and low AUCs in two sparse graphs, Power and Router whereas the GNN-based approaches are still able to provide high precisions and AUCs in both of the graphs. In both of USAir and Yeast graphs, SEAL shows the best results with precision of 0.94 and 0.89 and AUC of 0.96 and 0.98 respectively while the lowest precision and AUC values are recorded for WLNM and LLHN respectively. The use of node attributes for SEAL in USAir during prediction task influences in the improvement of the performance metrics. Overall, SEAL shows the highest AUC values in all graphs. The use of latent feature along with structural feature is the vital reason behind this success. Table~\ref{Table:small} shows that GNN-based approaches provide high-performance metrics in all graphs while similarity-based approaches perform well in some graphs.

The approaches are further evaluated in three large graphs FB15K, WN18 and YAGO3-10 and the results are presented in Table \ref{Table:large}.
We can see that some similarity-based approaches (AA, CN, PA, RA, IA, CAR) show higher metric values while others (JA, SA, SO, HDI, LLHN) show lower metric values than the GNN-based approaches in FB15K graph. The highest precision score is found for CN, IA, CAR approaches and the highest AUC value is found for SEAL. LLHN is the worst performing approach concerning both of the metrics among all approaches in FB15K graph. However, the precision drops to below 0.1 for all similarity-based approaches when applying the threshold to similarity scores with FB15K graph.
\renewcommand{\arraystretch}{1}
\begin{table}
\centering
\caption{AUC and Precision values with Max Scores and Min Scores in large graphs. Similar to Table ~\ref{Table:small}}/
\label{Table:large}
\scriptsize
\begin{tabular}{|p{1.5cm}|g{2cm}|g{2cm}|g{2cm}|g{2cm}|} 
\hline
\rowcolor{Gray}
\textbf{ App.}          & \textbf{Metrics}   &\textbf{FB15K} &\textbf{WN18} & \textbf{YAGO3-10} \\ \hline
\rowcolor{white}
\multirow{3}{*}{AA} & Precision &\textit{ 0.77}(0.0002)       &\textit{\textbf{0.13}}(0.0002)    &0.15(0.0018)  \\ \cline{2-5}
            & Max Score       &418.60  &57.32   &24.44  \\ \cline{2-5} 
            \rowcolor{white}
            & Min Score      &0.12    &0.00     &0.00  \\ \cline{2-5}
            & AUC       &\textit{0.82}        &0.56     &\textit{0.48}     \\ \hline 
\rowcolor{white}
\multirow{3}{*}{CN} & Precision &\textit{\textbf{0.81}}(0.0003)       &\textit{\textbf{0.13}}(0.0004)    &0.15(0.0012)   \\ \cline{2-5} 
            & Max Score      &1231.3  &60.00  &98.00   \\ \cline{2-5} 
            \rowcolor{white}
            & Min Score      &1.00     &0.00   &0.00   \\ \cline{2-5} 
            & AUC       &\textit{0.80}       &0.57    &\textit{0.48}   \\  \hline
\rowcolor{white}
\multirow{3}{*}{PA} & Precision &\textit{0.79}(0.0003)       &\textit{0.63}(0.0006)    &0.83(0.0006)   \\ \cline{2-5} 
            & Max Score      &9881842.3  &10636.7   &2426939  \\ \cline{2-5}
            \rowcolor{white}
            & Min Score      &942.67      &6.33       &109.00  \\ \cline{2-5} 
            & AUC       &\textit{0.88}       &\textit{0.64}    &0.88   \\ \hline
\rowcolor{white}
\multirow{3}{*}{RA} & Precision &\textit{0.77}(0.0003)       &\textit{\textbf{0.13}}(0.0002)    &0.15(0.0011)   \\ \cline{2-5} 
            & Max Score      &72.06  &20.67   &5.16  \\ \cline{2-5} 
            \rowcolor{white}
            & Min Score      &0.00   &0.00    &0.00  \\ \cline{2-5} 
            & AUC       &\textit{0.84}       &0.57    &0.57   \\ \hline
\rowcolor{white}
\multirow{3}{*}{JA} & Precision &\textit{\textbf{0.64}}(0.0225)       &\textit{\textbf{0.13}}(0.0161)    &\textbf{0.15}(0.0059)   \\ \cline{2-5} 
            & Max Score      &0.50  &0.50   &0.50  \\ \cline{2-5} 
            \rowcolor{white}
            & Min Score      &0.01  &0.00   &0.00  \\ \cline{2-5} 
            & AUC       &\textit{0.68}       &\textbf{0.56}    &\textit{0.46}   \\ \hline
\rowcolor{white}
\multirow{3}{*}{SA} & Precision &\textit{0.65}(0.0236)       &\textit{\textbf{0.13}}(0.0218)    &0.15(0.0068)   \\ \cline{2-5} 
            & Max Score      &1.00  &1.00   &1.00   \\ \cline{2-5} 
            \rowcolor{white}
            & Min Score      &0.02  &0.00   &0.00  \\ \cline{2-5} 
            & AUC       &\textit{0.70}       &0.57    &\textit{0.47}   \\ \hline
\rowcolor{white}
\multirow{3}{*}{SO} & Precision &\textit{\textbf{0.64}}(0.0225)       &\textit{\textbf{0.13}}(0.0180)    &0.15(0.0059)   \\ \cline{2-5} 
            & Max Score      &1.00  &1.00   &1.00  \\ \cline{2-5} 
            \rowcolor{white}
            & Min Score      &0.01  &0.00   &0.00  \\ \cline{2-5} 
            & AUC       &\textit{0.69}       &0.57    &\textit{0.46}   \\ \hline
\rowcolor{white}
\multirow{3}{*}{HPI} & Precision &\textit{0.69}(0.0959)      &\textit{\textbf{0.13}}(0.0796)    &0.15(0.0476)  \\ \cline{2-5} 
            & Max Score      &1.00  &1.00   &1.00   \\ \cline{2-5} 
            \rowcolor{white}
            & Min Score      &0.05  &0.00   &0.00   \\ \cline{2-5} 
            & AUC       &\textit{0.75}       &0.56    &\textit{0.47}   \\ \hline 
\rowcolor{white}
\multirow{3}{*}{HDI} & Precision& \textit{\textbf{0.64}}(0.0137)      &\textit{\textbf{0.13}}(0.0121)    &0.15(0.0035)  \\ \cline{2-5} 
            & Max Score      &1.00  &1.00   &1.00   \\ \cline{2-5} 
            \rowcolor{white}
            & Min Score      &0.01  &0.00   &0.00   \\ \cline{2-5} 
            & AUC       &\textit{0.68}       &0.57    &\textit{0.46}   \\ \hline
\rowcolor{white}
\multirow{3}{*}{LLHN} & Precision &\textit{\textbf{0.64}}(0.0008)     &\textit{\textbf{0.13}}(0.0046)    &0.15(0.0003)   \\ \cline{2-5}
            & Max Score      &0.28  &1.00   &1.00   \\ \cline{2-5} 
            \rowcolor{white}
            & Min Score      &0.00  &0.00   &0.00   \\ \cline{2-5} 
            & AUC       &\textbf{0.57}       &0.57    &\textit{\textbf{0.45}}  \\  \hline
\rowcolor{white}
\multirow{3}{*}{IA} & Precision &\textit{\textbf{0.81}}(0.0003)       &\textit{\textbf{0.13}}(0.0505)   &0.15(0.0014)   \\ \cline{2-5}
            & Max Score      &757.1  &4.58   &95.23   \\ \cline{2-5} 
            \rowcolor{white}
            & Min Score      &2.00    &0.00   &0.00   \\ \cline{2-5} 
            & AUC       &\textit{0.80}       &0.57    &\textit{0.47}   \\ \hline 
\rowcolor{white}
\multirow{3}{*}{CAR} & Precision &\textit{\textbf{0.81}}(0.0003)       &\textit{0.13}(0.0004)    &0.15(0.0008)   \\ \cline{2-5} 
            & Max Score      &6906  &60.00   &1430   \\ \cline{2-5} 
            \rowcolor{white}
            & Min Score      &1.00     &0.00    &0.00   \\ \cline{2-5} 
            & AUC       &\textit{0.80}       &0.57    &\textit{0.48}  \\ \hline
\rowcolor{white}
\multirow{3}{*}{CCLP} & Precision &\textit{0.78}(0.0015)       &\textit{\textbf{0.08}}(0.0006)    &0.14(0.0013)   \\ \cline{2-5} 
            & Max Score      &51.74  &1.67   &20.77   \\ \cline{2-5} 
            \rowcolor{white}
            & Min Score      &0.01   &0.00   &0.00   \\ \cline{2-5} 
            & AUC       &\textit{0.84}       &\textit{0.54}    &0.57  \\ \hline 
\rowcolor{white}
\multirow{3}{*}{WLNM} & Precision &\textit{0.67}     &\textit{\textbf{0.84}}    &0.68   \\ \cline{2-5}
            & AUC       &\textit{0.68}       &\textit{0.79}    &0.72     \\ \hline
\rowcolor{white}
\multirow{3}{*}{SEAL} & Precision &0.77 &\textit{0.61}    & \textit{\textbf{0.86}}   \\ \cline{2-5}
            & AUC       &\textbf{0.96} &\textit{\textbf{0.87}}    &\textit{\textbf{0.97}}    \\ \hline
\end{tabular}
\end{table}
 
 As shown in Table~\ref{stTable}, WN18 is a sparse graph with low average node degree (3.709) and clustering coefficient (0.077). This sparsity affects the performance of similarity-based approaches as these approaches except PA highly depend on the common neighbourhood information. The precision scores of all similarity-based approaches are below 0.2 except PA that shows a comparatively good precision score of 0.63. The precision further drops when applying the threshold to similarity scores in top-L links. Compared to the similarity-based approaches, GNN-based approaches show higher precision and AUC values in WN18 graph. The highest precision and AUC values are recorded for WLNM and SEAL approaches respectively. In YAGO3-10 graph, PA performs surprisingly well with  precision and AUC values of 0.83 and 0.88 respectively. However, the highest precision and AUC values are found for the SEAL approach. Overall, GNN-based approaches are more suitable across graphs from several domains with respect to precision and AUC values. 
 
From Tables~\ref{Table:small} and~\ref{Table:large}, the node-degree based approach PA shows higher performance comparing to other neighborhood based similarity approaches. The highest precision of PA is found in USAir (0.92) and the lowest one in Router(0.41). Similarity-based approaches based on the common neighborhood show impressive performance in the graphs with high average node degree and clustering coefficient. These approaches show very high precision of above or nearly 0.9 in two well connected Ecoli and USAir graphs. These approaches show very low precision of less than 0.2 in two large graphs, WN18 and YAGO3-10. On the other hand, the GNN-based approaches show very high precision and AUC across all of the experimental graphs including small to large graphs.

\subsection{Comparison of Computational Time}
The performance is further described in terms of computational time. Every approach is executed for each test set of each graph and their computational times are recorded. The computational time for similarity-based heuristic is the average time required per test link to compute the nodes similarity score. On the other hand, the computational times for GNN-based prediction approaches are the accumulated time for training the GNN and predicting the classes of links (existence or non-existence) in test sets. Table~\ref{Table:compTable} shows the mean computational time in milliseconds.
\renewcommand{\arraystretch}{1}
\begin{table}
\centering
\caption{Computational time (milliseconds). The graph-wise highest and lowest mean computational time are indicated in bold fonts and approach-wise highest and lowest mean computational time are indicated in italic.}
\scriptsize
\label{Table:compTable}
\begin{tabular}{|p{1.4cm}|p{0.8cm}|p{1.2cm}|p{0.6cm}|p{0.6cm}|p{1cm}|p{1.2cm}|p{1cm}|p{1cm}|p{1.1cm}|p{1.05cm}|p{0.5cm}|} 
\hline
\rowcolor{Gray}
\textbf{Approach} &\textbf{ Ecoli}	&\textbf{FB15K}	&\textbf{NS}	&\textbf{PB}	&\textbf{Power}	&\textbf{Router}	&\textbf{USAir}	&\textbf{WN18}	&\textbf{YAGO 3-10}	&\textbf{Yeast}	\\ \hline
AA  &221    &495    &71     &106    &73     &74     &\textit{15}     &288    & \textit{910}    &107		\\ \hline
JA  &28     &121    &26     &25     &63     &60     &\textit{13}     &256	&\textit{647}    &27		\\ \hline
PA  &\textbf{28}    &\textbf{120}   &\textbf{21}    &\textbf{23}    &61	&58 & \textit{\textbf{12}}    &251    &\textit{642}    &26 \\ \hline
RA  &330    &494    &70     &110    &65 	&\textit{63}     &104	&274	&\textit{915}    &95 \\ \hline
CN  &30     &120	&24     &24     &59     &56     &14     &249	& \textit{\textbf{629}}   &\textbf{25}    \\ \hline
SA  &58	    &226	&40     &48     &105	&102	&\textit{21}     &480	&\textit{1310}   &48 \\ \hline
SO  &60	    &228	&44     &47     &104	&98     &\textit{22}	    &476	&\textit{1298}   &49 \\ \hline
HPI &87	    &236	&63	    &70	    &149	&142	&\textit{33}	    &493	&\textit{1400}   &70 \\ \hline
HDI &60	    &227	&40	    &49	    &102	&96	    &\textit{19}	    &466	&\textit{1367}   &47 \\ \hline
LLHN &63	&147	&39	    &48	    &99	    &95	    &\textit{20}     &465	&\textit{1370}   &47 \\ \hline
IA  &412	&420	&57	    &218	&57	    &\textit{55}	    &82	    &262	&\textit{938}    &103    \\ \hline
CAR &280	&303	&54	    &134	&\textbf{54}	&\textit{\textbf{53}}	&63	&\textbf{157}	&\textit{643}    &117    \\\hline
CCLP &409	&654	&108	&280	&157	&\textit{152} 	&257	&492    &\textit{1696}   &255    \\ \hline
WLNM &612	&837	&170	&453	&257	&245	&\textit{153}	&541	&\textit{1440}   &363    \\ \hline
SEAL &\textbf{886}	&\textbf{1221}	&\textbf{398}	&\textbf{940}	&\textit{\textbf{340}}	&\textbf{419}	&\textbf{524}	&\textbf{868}	& \textit{\textbf{2713}}  &\textbf{403}   \\ 
\hline
\end{tabular}
\end{table}
From Table~\ref{Table:compTable}, it is seen that PA has the lowest mean computational time among the similarity-based approaches in half of the graph sets as it requires a simple multiplication operation of degrees of two end nodes in a link. The computational time for simple CN approaches are close to the PA approaches in all approaches. Similarity approaches those quantify the role of each neighbour or level-2 links such as JA, RA, IA require higher processing time. The highest computational times are found for CCLP similarity-based approach in all graphs as CCLP explores level-3 links for computing the similarity score. However, CAR requires the minimum computational time to predict links in sparse graphs (Power, Router, WN18) as these graphs have very lower clustering coefficient comparing to other graphs. The computational times of these approaches are affected by the graph properties such as average node degree, number of nodes and links, average clustering coefficient. For example, the computational time of all similarity-based approaches in NS graph is more than in USAir as NS is larger than USAir in terms of the number of nodes and link. The computational time in PB graph is more than in NS approaches as PB has more average node degree than NS though the number of nodes is higher in NS. \\
Compared to similarity-based approaches, the computational times of GNN-based ones are higher as they learn the heuristics from the graph during the training operation. Table~\ref{Table:compTable} shows that the computational times for SEAL are greater than WLNM in all graphs as SEAL utilizes the structural, latent and explicit features of graph comparing whereas WLNM utilizes only the structural features of the graph. One noticeable point is that the computational time of WLNM is more in PB, NS graphs than USAir as USAir is the smallest graph whereas SEAL reverses the case as it uses the node attributes in USAir. The highest computational time is recorded for SEAL among all the studied approaches. We also see that the computational times for GNN-based approaches grow by more amount than the similarity-based approaches. For example, the minimum computational time for PA in USAir grows by an amount of 629 milliseconds in YAGO3-10 graph whereas for SEAL it grows by an amount of 2189 milliseconds. Overall, similarity-based approaches are more efficient than GNN-based approaches concerning the computational time. Except for SEAL, the approach-wise comparison in terms of computational time shows that all approaches show the highest and lowest computational time in the largest experimental graph (YAGO3-10) and smallest graph (USAir) respectively, as expected. SEAL shows higher computational time in USAir than two sparse graphs (Router and Power) as it uses the attribute features of USAir and also the latter graphs have low average node degree.

\section{Conclusion}
In this paper, we study several link prediction approaches for homogeneous graphs from similarity-based and GNN-based learning categories with their working principles and limitations. The approaches were evaluated against ten benchmark graphs with different properties from various domains. The precision of similarity-based approaches was computed in two different ways to overcome the difficulty of tuning the threshold for deciding the link existence based on the similarity score. 

The experimental results show the superiority of GNN-based approaches over similarity-based ones with respect to the prediction performance across various graphs. In contrast, compared to similarity-based approaches, these GNN-based approaches are less suitable when the graphs need fast processing. The computational time of GNN-based approaches is further affected when applied to large graphs. In addition, the 'black box' problem of conventional neural networks remains unsolved with GNNs where it is very difficult to retrace the internal process of GNN. This work could help a new user to study similarity and GNN-based link prediction approaches and also the corresponding evaluation protocols.  

One perspective of this work is to achieve a good trade-off between prediction accuracy and computational time by developing a GNN-based link approach in a distributed and parallel environment. In addition, the approach is expected to be applicable to the heterogeneous graphs such as knowledge graphs.

%
%
%
%

\end{document}